\documentclass[aps,pre,twocolumn,showpacs,amssymb]{revtex4}
\usepackage[dvips]{graphicx}
\usepackage{dcolumn}% Align table columns on decimal point
\usepackage{bm}% bold math
\def\be{\begin{equation}}
\def\en{\end{equation}}

\def\gs{\gtrsim}
\def\ls{\lesssim}
\newcommand{\bi}[1]{\mbox{\boldmath$#1$}}

\def\bea{\begin{eqnarray}}
\def\ena{\end{eqnarray}}

\begin{document}

%\preprint{}

\title{Transitions among crystal, glass, and liquid 
 in a binary mixture  with changing  particle size ratio 
and temperature }


\author{Toshiyuki Hamanaka}
%\email[]{hamanaka@scphys.kyoto-u.ac.jp}
%\thanks{}
\affiliation{Department of Physics, Kyoto University, Kyoto 606-8502,
Japan}

\author{Akira Onuki}
%\email[]{onuki@scphys.kyoto-u.ac.jp}
%\thanks{}
\affiliation{Department of Physics, Kyoto University, Kyoto 606-8502,
Japan}


\date{\today}

\begin{abstract}
Using molecular dynamics simulation 
we examine changeovers  among  
crystal,  glass, and liquid 
 at high density   in  a 
two dimensional  binary mixture.  We  change  the ratio 
between the diameters of the two components 
and the temperature.  The transitions from crystal 
to glass or  liquid 
occur with proliferation of defects. 
We visualize the defects 
in terms of    a disorder variable    
$D_j(t)$ representing  a deviation 
from the hexagonal order for  particle $j$. 
The defect structures  are  heterogeneous 
and  are particularly  extended in polycrystal states. 
They look similar at the crystal-glass  crossover and 
at the melting.   Taking the average  of $D_j(t)$ over 
the particles,  we define  a disorder 
parameter  $D(t)$,  which   
conveniently  measures the degree of overall disorder. 
Its relaxation after quenching becomes   slow at low 
temperature in the presence of    size dispersity. 
Its steady state average   
 is small in  crystal and large  in  
glass and  liquid.   
\end{abstract}

%\keywords{}

\maketitle


\setcounter{equation}{0}

\section{Introduction}

The phase behavior of binary particle systems  
is much more complicated than that of 
 one component systems, 
where   the temperature $T$, the average number density 
$n=N/V$, and the composition are natural 
 control parameters. At high densities, 
 it  is  known to be profoundly influenced also 
by the size ratio $\sigma_1/\sigma_2$  between 
 the diameters of the two components, $\sigma_1$ and  $\sigma_2$  
\cite{Dick,Madden,Likos,Ito,Stanley}. 
If $\sigma_1/\sigma_2$  
is close to unity at  large $n$, 
the system becomes a crystal at low $T$ 
or a liquid at high   $T$.  
If $\sigma_1/\sigma_2$  considerably deviates from unity, 
glass states are  realized    
at  large $n$  and at low $T$. 
In glass states,   
the   particle motions  are nearly frozen  and the structural 
relaxation time grows, 
but the particle configurations are random 
yielding the structure factors similar to those in liquid.

Recently,  the liquid-glass transition    has   
been studied in  a large number of 
molecular dynamics simulations 
on model binary mixtures both in two and three dimensions  
\cite{Muranaka,Hurley,yo,kob,Barrat}.   
In these simulations, the temperature $T$ has 
mostly  been the control parameter 
at fixed average density and composition. 
Some authors have applied 
a  shear rate or  a  stress to glassy 
systems as a new control parameter 
\cite{yo,Barrat}. The size ratio $\sigma_1/\sigma_2$ 
has been chosen  at particular values 
to realize fully 
frustrated particle configurations and to avoid 
crystallization and phase separation. However, 
for weaker  size dispersity, 
the degree of disorder should become smaller. 
Polycrystals will be realized    
 at some stage and    a crystal  with 
a small number of point defects 
will  be reached eventually.  On this 
crossover  we  are not aware of any systematic study 
and have no clear picture.

In this paper,  we first   aim   to 
visualize the disorder brought about by  
the size dispersity  in two dimensions (2D).  
To this end we will introduce a disorder variable $D_j(t) \geq 0$ 
representing  a deviation of the  hexagonal crystal order 
  around each particle $j$. 
Snapshots of $D_j(t)$ realized by each simulation run  
 will  exhibit patterns indicating the nature of 
the defect structure. 
We shall observe point defects in crystal, grain boundaries 
in polycrystal, and amorphous  disorder in glass.  
The  average  $D(t)=\sum_{j=1}^N D_j(t)/N$ over the particles 
is   a single "disorder parameter"  characterizing 
the degree of overall disorder.

Halperin and Nelson \cite{NelsonTEXT} found  
that defects    play a key role 
in 2D  melting  in 
one component systems, predicting  
continuous transitions with an intermediate 
"hexatic" phase between crystal and liquid. 
They introduced a  sixfold orientation 
order variable, written as $\chi_j$  in this paper. 
The correlation function $g_6(r)$  of the thermal 
fluctuations of $\chi_j$ 
 has been used to characterize the 
2D  defect-mediated melting 
theoretically 
\cite{Frenkel,Stanley,Ito} 
and experimentally \cite{Rice,Maret,Kramer,quinn}.  
Our disorder variable $D_j(t)$ will be constructed 
from  their $\chi_j$, so we will  visualize the defect patterns 
exhibited by  $D_j(t)$ also at the melting. 
The  problem becomes much more complex  for binary 
mixtures, where the 
 crystal-liquid transition  occurs 
  with changing $T$ or $n$ 
at weak  size dispersity \cite{Stanley,Ito} 
and  the glass-liquid transition 
occurs at stronger  size dispersity 
\cite{Muranaka,Hurley,yo,kob,Barrat}.   
 We should 
 understand the  defect structure  
by  changing $\sigma_1/\sigma_2$   and $T$ (and/or $n$) 
  both at 
 the crystal-glass and  crystal-liquid transitions \cite{cg}.

In Sec II, we will introduce the quantities mentioned above 
and present our numerical results 
 at fixed density and composition, where   
the defects   involved in 
the crystal-glass and crystal-liquid transitions 
will be visualized.  
We will  also 
calculate the overall disorder parameter $D(t)$ 
in transient states and in steady states 
as a function of   $\sigma_1/\sigma_2$   and $T$.   
In Sec III, we  will summarize 
our results and give some remarks.

\section{Numerical results}
\subsection{Method}

We used  a 2D  model 
binary mixture interacting 
via a  truncated Lenard-Jones (LJ) potential  $v_{\alpha\beta} (r)$, 
where $\alpha, \beta=1,2$ represent the particle species. 
If the distance $r$ between two particles 
is larger than a  cut-off $r_{\rm cut}$, 
we set $v_{\alpha\beta} (r)=0$.  If $r<r_{\rm cut}$, 
it  is given by the Lennard-Jones potential, 
\be
v_{\alpha\beta} (r)=
4\epsilon \left[ \left(\frac{\sigma_{\alpha\beta}}{r}\right)^{12}
-\left(\frac{\sigma_{\alpha\beta}}{r}\right)^{6}\right] -C_{\alpha\beta} , 
\end{equation}
which  is  characterized by 
the energy  $\epsilon$   and 
the soft-core diameter $\sigma_{\alpha\beta}
=(\sigma_{\alpha}+\sigma_{\beta})/2$ 
with $\sigma_1$ and $\sigma_2$ representing  
 the (soft-core) diameters  
of  the two components. 
The constant $C_{\alpha\beta}$  ensures  
$v_{\alpha\beta} (r) \rightarrow 0$  as $r \rightarrow r_{\rm cut}$, 
so the potential is continuous at the cut-off distance. 
We set  $r_{\rm cut}=
3.2\sigma_{1}$   for any $\alpha$ and $\beta$ \cite{note2}. 
The particle numbers of the two species  are 
$N_{1}=N_{2}=500$, so $N=N_1+N_2=10^3$. 
With varying the size ratio $\sigma_2/\sigma_1$, 
the system  volume $V$ was 
changed such that the volume fraction 
of the soft-core regions defined by 
\be 
\phi=  (N_1\sigma_1^2+N_2\sigma_2^2)/V  
\en  
was fixed at $0.9$ mostly. We set  $\phi=1$ 
only in one case  (in 
the lower panel of Fig.8).  With the  mass ratio being  
 $m_{1}/m_{2}=(\sigma_{1}/\sigma_{2})^{2}$, we   integrated  the Newton   
equations using  the leapfrog algorithm   under the periodic boundary 
condition. The system temperature was controlled with 
 the  Nose-Hoover thermostat 
\cite{allen,frenkelbook,nose}.      The   time step 
of integration was  $0.002\tau$, where 
\be 
\tau=\sigma_{1}\sqrt{m_{1}/\epsilon}.
\en  
Hereafter the time $t$ and the temperature 
$T$ will be measured in units of $\tau$ and $\epsilon/k_B$, respectively.

We first equilibrated  
the system  in a liquid state at $T=2$ in a  
  time interval of  $10^3$ and then 
 quenched it to a lower final  temperature 
with further equilibration in a period of    
$ t_{\rm eq}= 1.1\times 10^4$ \cite{barker,hamanaka}. 
There was no appreciable time evolution 
in the pressure,  the energy, {\it{etc}} 
in the time region  $t \gs 4\times 10^3$ 
(see Fig.8 as an example)  \cite{equi}.  
The particles  were  well mixed and
  no indication of phase separation was observed 
in the final time region.

In our study, 
the  size ratio was   in the range  
 $1 \leq \sigma_1/\sigma_2\leq 1.4$. 
We  saw no  tendency of phase separation. 
If $\sigma_{1}/\sigma_{2}$ is  too large,   
 phase separation 
 will be detected  \cite{HansenS,Hobbie}.
We show  typical particle configurations in Fig.1 
at the final simulation time  $ t= 1.2\times 10^4$ 
  for (a) $\sigma_2/\sigma_1=1.1$, 
(b) $1.2$, (c) $1.225$, 
and (d) $1.4$  at $\phi=0.9$. 
The system  length $V^{1/2}$ is (a)  35.03,  
(b) 36.81,  (c) 37.27,  and (d) 40.55  in units of $\sigma_1$.
They  represent (a) 
 a crystal  state with point defects, 
(b) and (c)    polycrystal states, 
and  (d) a glass  state.

% Fig1

\begin{figure}[t]
\includegraphics[scale=0.4]{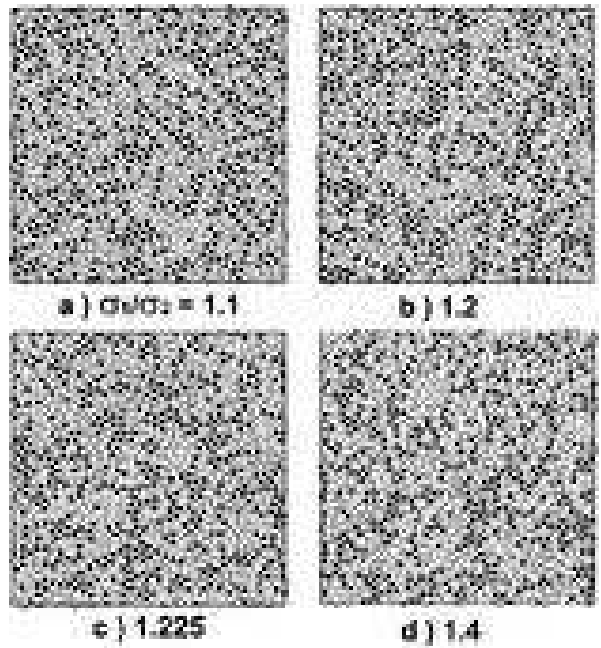}
\caption{Particle configurations for 
(a) $\sigma_{1}/\sigma_2=1.1$,  (b) 1.2, 
 (c) 1.225, and (d) 1.4 at  $\phi=0.9$  and $T=0.2$ 
 taken at the final simulation time 
 $ t= 1.2\times 10^4$.  
Smaller (larger) circles represent  
the smaller (larger) particles. 
For the visualization purpose 
the diameters of the circles in the snapshots are 
taken as $A\sigma_1$ and $A\sigma_2$ with $A<1$ 
for the two species.  
}
\end{figure}

\subsection{Sixfold orientation order}

In Fig.1,  a large fraction of 
the particles are enclosed by six particles 
even at $\sigma_1/\sigma_2= 1.4$. 
The  particle  configurations 
are remote from other ordered structures   
such as the square structure \cite{Likos}. 
Therefore, we  consider 
deviations from the hexagonal order.  
The  local crystalline order 
is  represented  by a  sixfold 
orientation order variable  
\cite{NelsonTEXT}. 
 For each particle $j$ we define 
\be 
\chi_j  = \sum_{k\in {\rm bonded}}\exp[6i\theta_{jk}],  
\en   
where the summation is over  the particles  
''bonded''  to the particle $j$.  
In our case, 
 the two particles $j \in\alpha$ and 
$k \in\beta$ are bonded,  if  their  distance 
$r_{jk}= |{\bi r}_j-{\bi r}_k|$ is shorter than 
$R_{\alpha\beta}=1.25\sigma_{\alpha\beta}$ \cite{yo}. 
The upper cut-off $R_{\alpha\beta}$  is slightly 
longer than the first peak position of 
the pair-correlation function $g_{\alpha\beta}(r)$.  
The  $\theta_{jk}$ is the angle of the relative vector 
${\bi r}_j-{\bi r}_k$ with respect to the $x$ axis. 
For  a perfect triangular crystal of a one component system,  
the complex numbers   $\chi_j$  are all equal to 
$6\exp(6i\alpha)$  
with $\alpha$ being the common angle of one of the crystal axes 
with respect to the $x$ axis. In the presence of disorder,  
the absolute values $|\chi_j|$  are significantly 
different from 6 for particles around defects. 
It is  convenient to 
define a local crystalline angle $\alpha_j$ 
in the range  $0\le \alpha_j<\pi/3$  by 
\be 
\Phi_{j}=\chi_{j}/|\chi_{j}|=e^{6i\alpha_{j}}.
\label{DefinitionOfPhi}
\en

In Fig.2, we show the snapshots of the angles 
 $\alpha_j$ ($j=1, \cdots, 10^3$) for the same 
particle configurations in Fig.1. 
The color map is illustrated in Fig.3. 
We   can clearly see  point defects, 
  grain boundaries, 
and    glassy particle configurations.  In (b) 
the grain boundaries are localized, while in (c) they are 
percolated.   Recently, using a 2D model of block copolymers, 
  Vega {\it et al.} \cite{Vega} numerically  studied 
the  grain boundary coarsening   
to obtain pictures of the orientation angles 
similar to our Fig.2, though their system 
corresponds to one component particle systems.

%%%%%%%%%%%%%%% Fig2
\begin{figure}[t]
\includegraphics[scale=0.4]{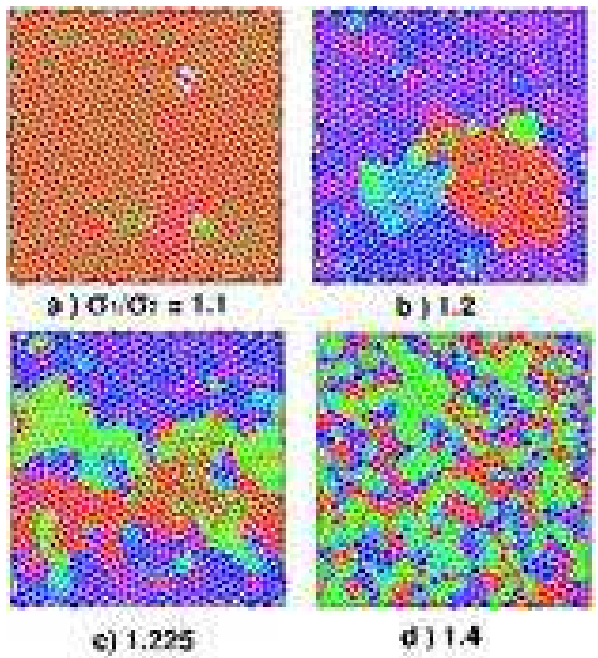}
\caption{Snapshots of the angles   $\alpha_j$ in Eq.(5) for 
(a) $\sigma_{1}/\sigma_2=1.1$, (b) 1.2,  (c) 1.225,  
 and (d) 1.4    with the  the  color map 
in Fig.3.  The data are common to those in Fig.1.
Changeover  from  crystal 
to  glass  occurs with  polycrystal  
 as an intermediate state. 
}
\label{rot}
\end{figure}

% Fig3
\begin{figure}[h]
\includegraphics[scale=0.2,angle=90]{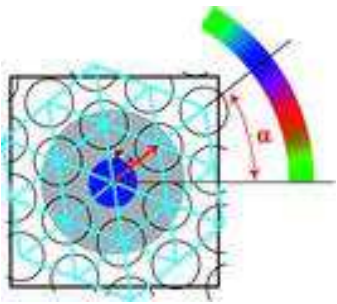}
\caption{
Color map for the angle $\alpha_j$ in Eq.(5) for  particle $j$ 
at the center, around which the 
crystal order  is perfect. The gray circle is the bonded  region. 
The vector ${\bi r}_{jk}$  (red arrow) 
makes an angle of $\alpha_j = 40^{\circ}$ with respect to the 
horizontal axis. The color 
of particle $j$  
is  then blue.}
\label{lattice}
\end{figure}

\subsection{Disorder variable}

% Fig4
\begin{figure}[h]
\includegraphics[scale=0.5]{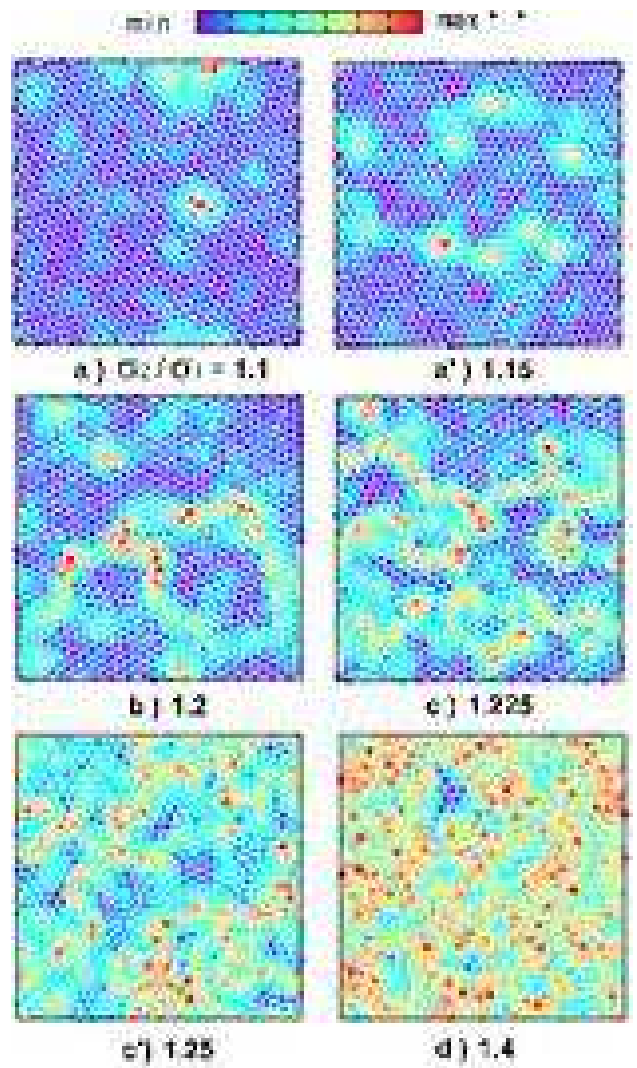}
\caption{Disorder variable  $D_{j}$  in Eq.(6) 
for   (a) $\sigma_2/\sigma_1=1.1$, (a') $1.15$, (b) $1.2$, 
(c) $1.225$,  (c') $1.25$, and (d) $1.4$ with 
 $\phi=0.9$ and  $ t= 1.2\times 10^4$. 
The particle  configurations in (a), (b), (c), and (d) 
are common to those in Figs.1 and 2.  
Here the color changes in the order of rainbow. 
}
\end{figure}
%Fig5
\begin{figure}[h]
\includegraphics[scale=0.3,angle=90]{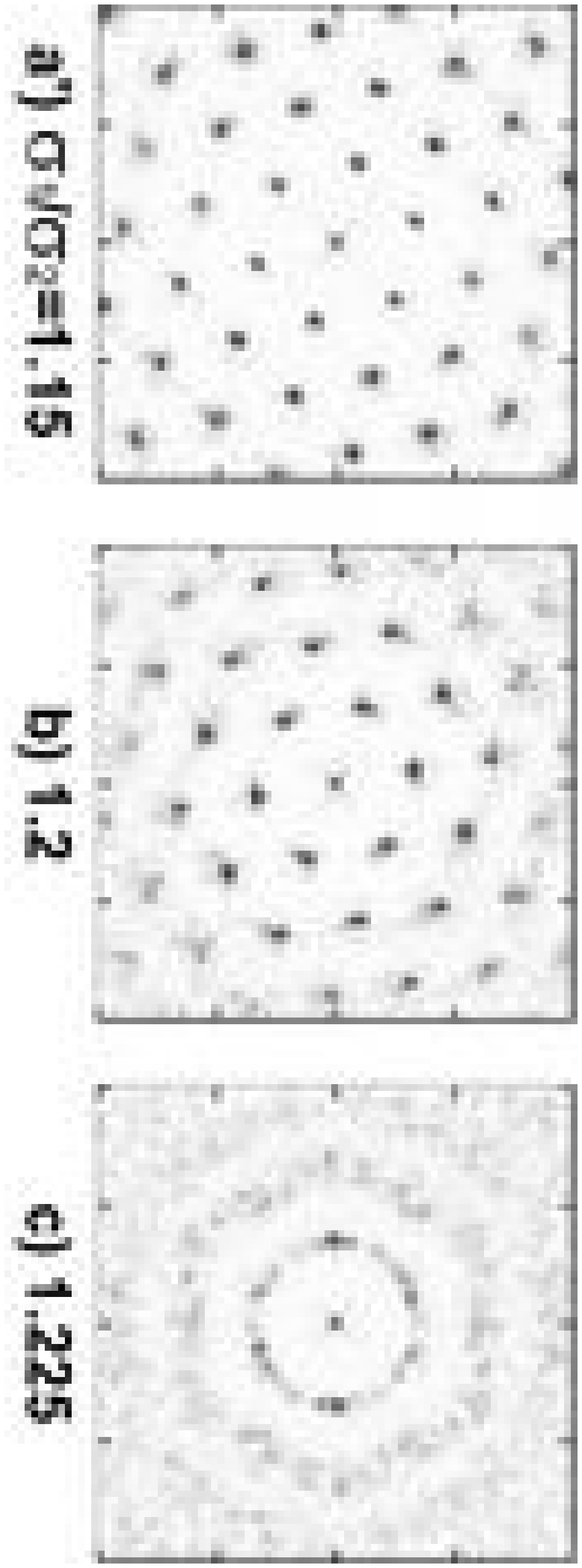}
\caption{Structure factor 
for   (a') $\sigma_2/\sigma_1=1.15$, (b) $1.2$, 
and (b') $1.225$ in Fig.4. Bragg peaks can be seen 
in  (a) and (b), while it resembles to that in liquid 
for (c) (see (c) in Fig.7). 
}
\end{figure}
%Fig6
\begin{figure}[h]
\includegraphics[scale=0.3,angle=-90]{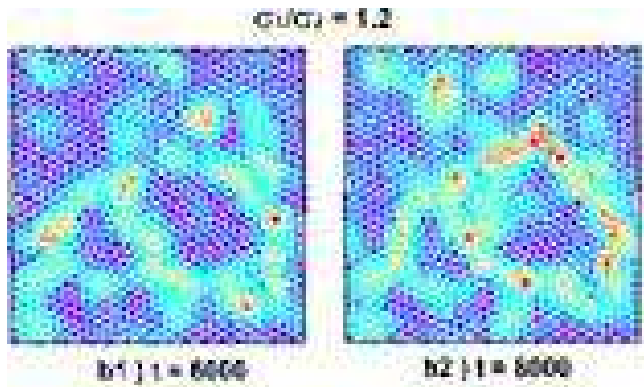}
\caption{Disorder variable   $D_j$  in Eq.(6) 
in a polycrystal state 
at $\sigma_2/\sigma_1=1.225$ for 
(b1) $t=6\times 10^3$ 
and (b2) $t=8\times 10^3$,  
in the same run 
giving the panel (b) in Fig.4 at $t=12\times 10^3$.  
Comparison of these three snapshots indicates  very slow 
time evolution of the grain boundaries. See points (b1), (b2), and (b) 
in Fig.7 also. 
}
\end{figure}

We next introduce a new  variable representing 
the degree of disorder. 
In  terms of the difference 
$\Phi_{k}-\Phi_{j}$ between the bonded 
particle pairs,  we  define  
\bea
D_{j} &=&{\sum_{k\in bonded}
|\Phi_{j}-\Phi_k|^{2}}\nonumber\\
&=&  2\sum_{k\in bonded}
 [1-\cos 6(\alpha_j-\alpha_k)], 
\ena
for each particle $j$. This quantity  is  called 
the disorder variable.  
If the thermal vibrations are neglected,  
$D_{j}$ vanishes in single-component
 perfect  crystals and 
is  nonvanishing around defects. 
It takes large values of order unity 
almost everywhere in highly frustrated glass states. 
See the comment  (iv) in the last section 
for  appropriateness 
of  this variable in glass and liquid.

In Fig.4,  snapshots of $D_{j}$ are shown for 
  (a) $\sigma_2/\sigma_1=1.1$, (a') $1.15$, (b) $1.2$, 
(c) $1.225$,  (c') $1.25$, and (d) $1.4$   at  
the final  states of the simulation runs at  $ t= 1.2\times 10^4$. 
 Those of  (a), (b), (c), and (d) 
are taken from the same particle configurations 
 as in  the corresponding panels of  Figs.1 and 2.  
The color of the particles varies 
 in the order of rainbow, being 
 violet   for  $D_{j}=0$ and  
red for  the  maximum of $D_j$. 
In Fig.4, the maximum of $D_j$ is  
1.71, 2.64, 3.59, 4.36, 4.34, and 4.58 
in (a)-(d) in this order.  
In  crystals with  $\sigma_2/\sigma_1$ close to unity, 
a small number of defects 
can be detected   as  bright points as in (a) and (a'). 
In polycrystals,  defects are accumulated 
to form  grain boundaries 
detectable as bright  closed curves  
enclosing small crystalline  regions, as in (b) and (c). With 
further increasing  $\sigma_2/\sigma_1$,  
defects are proliferated and a large 
 fraction of the particles 
are depicted  as  bright points. 
In the largest size ratio in (d),  
 most of the particles are in disordered 
configurations.   With varying $\sigma_2/\sigma_1$,   
this crossover occurs  
in a narrow range around 1.2.

In Fig.5,  the structure factor 
of the number density 
$n({\bi r}) =n_1({\bi r})+n_2({\bi r})$ 
is written for (a) $\sigma_2/\sigma_1=1.15$, 
(b) 1.2, and (c) 1.225  
to confirm the abruptness of 
this crystal-glass crossover. 
The structure factor   in (a) 
exhibits Bragg peaks showing translational 
order,  while  that   in (c)  
 is similar to that in liquid  but still retains  
the sixfold angular symmetry. 
For the intermediate case (b), 
 the sixfold  symmetry is evidently present   and  
the translational order is being lost.  
(See Fig.9 below   for structure factors 
in  typical  cases far from the transitions.)  
We note that similar 
 structure factors 
were taken  from  a quasi 2D colloid 
suspension around the melting \cite{Rice}.

The timescale of the 
particle  configurations becomes exceedingly slow 
in glass states 
\cite{Muranaka,Hurley,yo,kob,Barrat}.   
 Also in polycrystal states, 
 the motions of the grain boundaries  
become  slow  with increasing $\sigma_2/\sigma_1$, 
while the grain boundaries coarsen  to disappear 
 in one component systems on a  rapid  timescale 
(see the corresponding curve in Fig.7) \cite{Vega}. 
In Fig.6, we present two additional snapshots 
of $D_j(t)$  at $\sigma_2/\sigma_1=1.225$ for 
$t=6\times 10^3$ and $t=8\times 10^3$, 
while   the panel (b) in Fig.4 is the snapshot 
at $t=12\times 10^3$  in the same run. 
These three snapshots exhibit percolated grain boundaries 
with only small differences 
on  large scales, indicating  
 pinning of the grain boundaries. 
The panel (b) in Fig.2 demonstrates   that 
the system is a polycrystal.  As a result, we cannot deduce 
the life time of the gran boundaries from our simulation  in this case.

%Fig7 
\begin{figure}[h]
\includegraphics[scale=0.3,angle=-90]{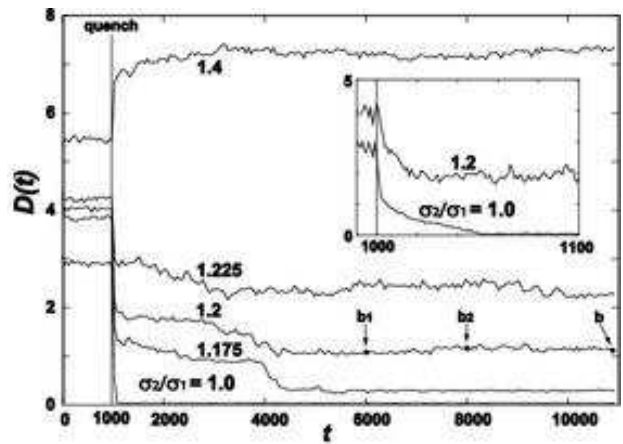}
\caption{Relaxation of the disorder 
 parameter  $D(t)$  in Eq.(7)  at $\phi=0.9$ 
for  $\sigma_2/\sigma_1=1.0$, $1.175$, $1.2$,  $1.225$, 
 and $1.4$  from below.  The temperature $T$ is 
lowered from 2 to 0.2 at $t=10^3$. 
For the one component case $\sigma_2/\sigma_1=1.0$,  
it takes place on a timescale of 50, as shown in the expanded 
 inset. With size dispersity,  
$D(t)$   relaxes  slowly on timescales 
of order $4\times 10^3$. Snapshots  of $D_j(t) $  at 
two points (b1) and (b2) on the curve of 
$\sigma_2/\sigma_1=1.2$  are given in Fig.6. 
}
\end{figure}

%Fig8 
\begin{figure}[h]
\includegraphics[scale=0.4]{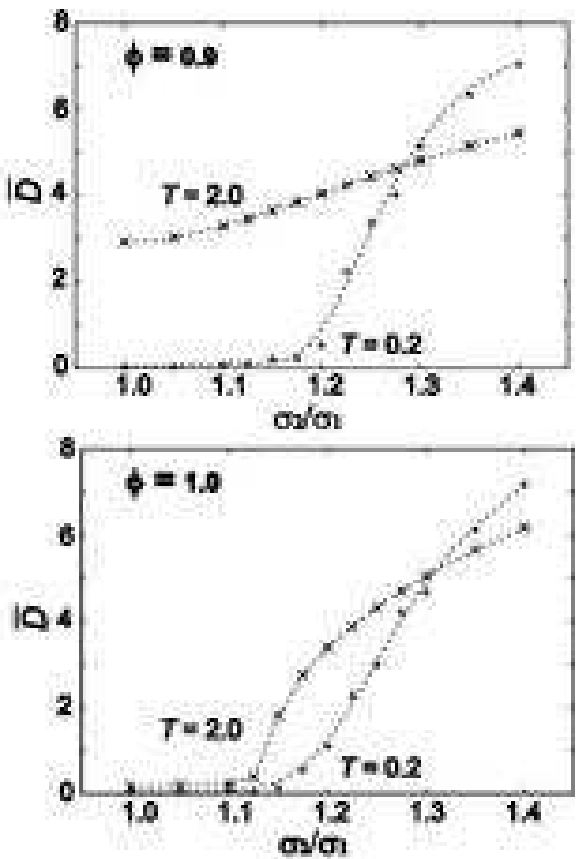}
\caption{Disorder parameter $\bar D$ in Eq.(8) 
for $T=0.2$ and $2$, with  $\phi=
0.9$ in the upper panel and 
$1$ in the lower panel. Liquid states are realized 
on  the curve of $T=2$ in the upper panel. 
The other curves show the crystal-glass crossover. 
}
\end{figure}

%Fig9
\begin{figure}[h]
\includegraphics[scale=0.4]{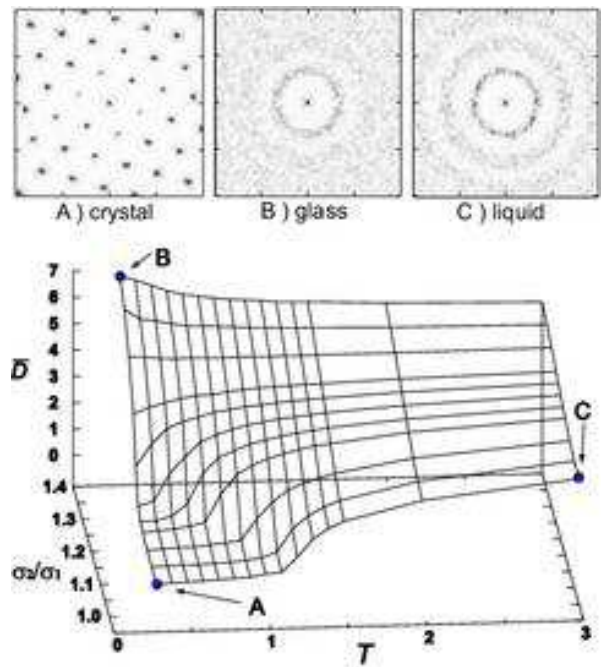}
\caption{Disorder parameter  $\bar D$  in Eq.(8) 
as a function of    $\sigma_2/\sigma_1$ 
and $T$ for $\phi=0.9$. Typical structure factors for 
(A)  crystal, (B) glass, and (C) liquid 
in the upper panel, where the  corresponding points are 
indicated in the lower panel. 
}
\end{figure}

%Fig10
\begin{figure}[h]
\includegraphics[scale=0.4]{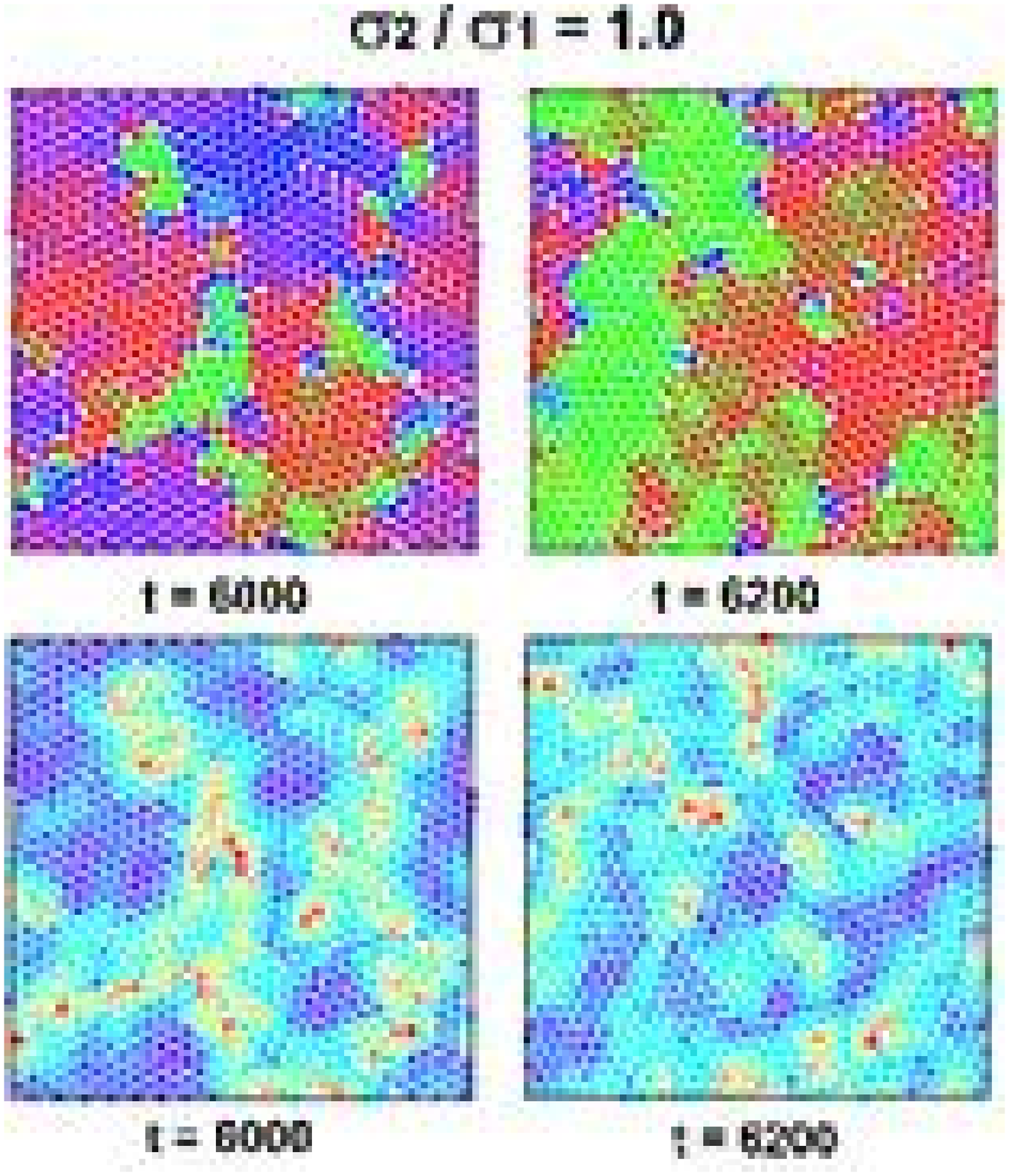}
\caption{Polycrystal configurations  of $\alpha_j$ in 
Eq.(5) (top)  and $D_j$ 
in Eq.(6) (bottom)   at $t=6\times 10^3$ (left) 
and $6.2\times 10^3$ (right) 
in the one component case 
  $\sigma_2/\sigma_1=1$ at $T=1.3$ 
and   $\phi=0.9$.  
The system is intermediate 
between crystal and liquid. The defect structure 
is evolving  rapidly on a timescale of 
$50$. }
\end{figure}

\subsection{Degree of overall disorder}

We  now introduce a single "disorder parameter" 
 representing  the   degree of overall disorder 
by taking the average over all the particles, 
\be  
D(t)=  \frac{1}{N}\sum_{j} D_{j}(t), 
\en
where the time-dependence of $D_j(t)$ and $D(t)$  
is explicitly written. 
  In Fig.7,  
we show time evolution of $D(t)$, where  quenching 
is from liquid at $t=10^3$.   It 
undergoes very slow time evolution 
with finite size dispersity in polycrystal and glass,  
in accord with Fig.6. For $\sigma_2/\sigma_1=1.4$,  
$D(t)$ increases upon quenching (see Fig.9 below for its reason). 
In the time region $t \gs 4\times 10^3$, 
we can see no appreciable relaxation in these curves. 
For the one component case, the relaxation from 
liquid to crystal terminates rapidly  on a timescale of 50.

However, the curves in Fig.7 with size dispersity 
weakly   depend  on time  around the 
average  even in apparent steady states.  
For example, $D(t)$  is 
1.10 in (b1) of  Fig.6,  1.17 in (b2) of  Fig.6, 
and 1.11 in (b) of Fig.4.  
 This temporal fluctuations 
should diminish  for larger   system size.  
Its deviation  from the time average 
became largest when  the grain boundaries appreciably moved 
in polycrystal  states.   
For each simulation run,  we  defined  the 
time average of $D(t)$ as   
\be  
{\bar D}=  \frac{1}{t_{\rm d}}\int_{t_{\rm f}-t_{\rm d}}^{t_{\rm f}} dt  D(t), 
\en 
where $t_{\rm f}$$(=1.2\times 10^4)$ 
 is the terminal  time of the simulation run  
and  $t_{\rm d}(=2\times 10^3)$  is the width of 
the  time interval of taking data. 
We regard $\bar D$ as a steady state average  
though glass states may further relax on  longer 
timescales.

In Fig.8,   we plot  $\bar D$ 
 as a function of  the size ratio. 
 In the upper panel, where  $\phi=0.9$, 
 liquid states are realized for 
any  $\sigma_2/\sigma_1$ at  $T=2$,  
while the system   is crystalline for  
 $\sigma_2/\sigma_1 \ls 1.2$ 
and glassy for larger  $\sigma_2/\sigma_1$  at 
$T=0.2$.  In the range  $\sigma_2/\sigma_1 \gs 1.3$, 
  $\bar D$  in the liquid state at  $T=2$ 
becomes smaller than $\bar D$ in the glass 
state  at  $T=0.2$.  This is because  
$\bar D$ increases weakly with increasing $\sigma_2/\sigma_1$ 
in liquid  and increases more strongly  
 in  glass.  In  the lower panel, where  $\phi=1$, 
the system crosses over from crystal  to 
glass    both for $T=0.2$ and 2,  and 
 $\bar D$ increases rather abruptly 
around   $\sigma_2/\sigma_1 \sim 1.2$. 
For   $\sigma_2/\sigma_1 \ls 1.1$,  
$\bar D$ takes a small positive number  
due to the thermal motions of the particles.   
In Fig.9,   $\bar D$  is plotted 
as a function of    $\sigma_2/\sigma_1$ 
and $T$.  It shows the overall behavior 
of $\bar D$.  That is,  $\bar D$   is small in crystal  and 
increases  abruptly in glass  and liquid.  
Interestingly, for $\sigma_2/\sigma_1 > 1.25$, 
$\bar D$ decreases with increasing $T$ 
from glass to liquid (see Fig.7). For such 
size ratios, highly 
 disordered particle configurations 
can be  pinned at low $T$  and the thermal 
motions at high $T$ can relax them.

\subsection{Defect-mediated melting}

In our  simulations  at fixed density, 
we observed 
defect proliferation  at the melting 
(as well as at the crystal-glass crossover)  and 
no coexistence of crystal and liquid regions separated by 
 sharp interfaces.  The system 
became   highly heterogeneous (as  in Fig.10 below), 
but  no nucleation process could be detected.  
   Figure 9 shows that  
$\bar D$ changes  continuously  
along the $T$ axis at each   
$\sigma_2/\sigma_1$ including  the one component limit 
 $\sigma_2/\sigma_1=1$.   Similarly, in a 2D  
Lenard-Jones system with $N=256$  at $\phi=0.8$, 
 Frenkel and Mctague detected 
no discontinuity in the average pressure 
and  energy  \cite{Frenkel}.  
Theoretically, the 2D melting  can be either continuous 
or first ordrer   depending the specific details of 
the system \cite{Saito,Chui}. It is   a delicate 
problem to determine its  precise nature  
in the presence of 
the heterogeneity developing at the transition 
\cite{Rice,Maret,Kramer,quinn}.

To visualize  the physical 
process involved at the melting, 
we display  snapshots of $\alpha_j(t)$ 
and $D_j(t)$ at $t=6000$ and 6200 in Fig.10  
  in the one component case at $T=1.3$, 
where the change of $\bar D$ is abrupt in Fig.9. 
The $D(t)$ in Eq.(7) is 
1.80 at $t=6000$ and 1.60 at $t=6200$. 
We can see percolated  grain-boundary patterns and chains of point defects. 
The area fraction of the crystalline regions 
with small $D_j$ continuously decreases (increases) 
with further raising (lowering) the temperature. 
We mention 
a simulation by  McTague {\it et al.} 
in a one component 
system with soft-disk  $r^{-6}$ potentials \cite{McTague},  
reporting the presence of both free dislocations 
and many  grain boundaries 
at the melting. Some authors already 
 pointed out relevance of 
grain boundaries in the 2D melting \cite{Fisher,Chui}. 
Using inherent-structures theory, 
Somer {\it et al.} \cite{Somer} 
found  percolated  grain boundaries   in 
"inherent structures" after 
 a hexatic-to- liquid transition. 
Among many experiments, grain boundaries 
were evidently shown   in Ref.\cite{quinn}.

We   notice close similarity   between 
 the snapshots  of the polycrystal states  
in Figs.6  and 10.  However, very different are 
 the timescales 
of the dynamics  of $D_j(t)$  
without  and with   size dispersity. 
Indeed, the patterns in Fig.10 changed  
appreciably  on a rapid timescale of 50, 
while   the large scale 
patterns in Fig.6 were  nearly frozen   
in our simulation time.

\section{Summary and remarks}

In summary, using  
MD simulation  on a 2D LJ binary mixture, 
we  have investigated the  effects  of 
the size dispersity in the range 
$1\leq \sigma_1/\sigma_2\leq 1.4$ 
and the temperature 
 in  the particle configurations at fixed average density 
and composition.  
Our main objective 
has been to  visualize  defects, so 
the system size ($N=10^3$) has been chosen 
to be  rather small. Larger system sizes 
are needed to get  reliable  correlation functions 
of the density and the  sixfold orientation variable.

 We summarize our main 
results and give  remarks.\\
(i)   
We have displayed the angle variable 
$\alpha_j$ defined by   Eq.(5) in Fig.2 and 
the disorder variable $D_j(t)$ 
defined by Eq.(6)  in Fig.4 at low $T=0.2$. 
The  snapshots of these variables 
 evidently show how the particle configurations 
become disordered with increasing the size ratio. 
Those  of $D_j(t)$ provide the real space pictures 
of   the defect structures 
on various spatial scales. 
We find polycrystal states with grain boundaries 
between crystal and glass.  
The motions of the grain boundaries are  much slowed down 
with size dispersity,  as  in Figs.6 and 7.\\
(ii)  
The disorder  parameter $D(t)$ in   Eq.(7) or 
its time average $\bar D$ in   Eq.(8) 
is  a  measure of 
 overall disorder. As in Fig.7, the relaxation of $D(t)$ 
after quenching from a high to a low 
temperature occurs on a very long timescale with 
 size dispersity, while 
it relaxes much faster in one component systems. 
The steady state average $\bar D$ 
is small in crystal and increases abruptly 
in glass and liquid,  with increasing  
$\sigma_1/\sigma_2$ or $T$,  as in Figs.8 and 9.\\ 
(iii)
In our system,  the crystal-glass and crystal-liquid 
crossovers  proceeded  with increasing the  
 defect density without nucleation.  
Remarkable resemblance 
is noteworthy between 
the polycrystal patterns of $D_j(t)$ at the crystal-glass transition 
in Figs.4 and 6 and  
those in the one component case 
at the crystal-liquid transition 
in Fig.10.  However, 
the timescale of the defect structure 
is drastically enlarged with increasing $\sigma_1/\sigma_2$.  
 In these two transitions,  
the disorder parameter $\bar D$ increases 
abruptly but continuously 
from  small (crystal) values  
to large (glass or liquid) values. 
In these  cases,  polycrystal states  appear 
 with large scale 
heterogeneities in $D_j(t)$, as can be seen in Figs.4, 6, and 10. 
\\
(iv) 
The particle configurations  will increasingly 
deviate from the hexagonal order in the crossover 
from crystal to glass or liquid. 
They might become  rather closer to other ordered 
structures for some fraction of the particles  \cite{Ito,Likos}. 
In such cases, large values of    
$D_j(t)$ and $\bar D$ will 
 have only qualitative meaning, 
since they represent deviations 
 from the hexagonal order.\\ 
(v) 
 We should  study the pinning mechanism 
of   grain boundaries in polycrystal 
 in the presence of size dispersity.  
We  should also examine the  dynamical properties 
such as the diffusion constant,  the shear viscosity, and 
the time-correlation functions  for various degrees of 
disorder.  They have been calculated 
around the liquid-glass transition 
\cite{Muranaka,Hurley,yo,kob,Barrat}.\\
(vi) For the pair potentials in Eq.(1) 
and for our limited simulation time, 
we have  detected no  tendency of phase separation. 
By increasing the repulsion among 
the different components, we could study  nucleation 
of crystalline domains in a glass matrix, for example.\\ 
(vii)   
We will  report shortly on the  shear flow effect   
 at the  crystal-glass and crystal-liquid 
transitions in 2D.  It has already been studied 
at the liquid-glass transition \cite{yo,Barrat}. 
It is of interest how an applied shear 
affects the defect structure and induces 
plastic deformations \cite{OnukiPRE}.

\begin{acknowledgments}
The calculations of this work 
were performed at the Human Genome Center, 
Institute of Medical Science, University of Tokyo. 
This work was  supported by 
Grants in Aid for Scientific 
Research 
and for the 21st Century COE project 
(Center for Diversity and Universality in Physics)
 from the Ministry of Education, 
Culture, Sports, Science and Technology of Japan.
\end{acknowledgments}


\begin{references}
\bibitem{cg} Experimentally, we need different systems 
to change the size ratio, so we should use 
"the crystal-galss crossover" rather than 
"the crystal-glass trsnsition", to be precise.
\bibitem{note2} 
The LJ potential among  the larger particles 
is  minimum at $r=1.12\sigma_{1}$, which 
 is shorter than 
the adopted value of $r_{\rm cut}$  
even for the maximum size ratio 
$1.4$.
\bibitem{equi}  Our simulation time 
becomes  shorter than the 
structural relaxation time 
in glass.    Furthermore, Fig.6 
shows that it is  shorter than the timescale of the grain boundaries 
at $\sigma_1/\sigma_2=1.2$. Much 
 longer simulation times are thus needed 
for more quantitative analysis  of such states.
\end{references}

\newpage

\end{document}